\documentclass[10pt]{article}

\usepackage{amsfonts}
\usepackage{amsmath}
\usepackage{amssymb}

\usepackage{enumerate}
\usepackage{graphicx}
\usepackage{bm}
\usepackage{color} 
\usepackage{multicol}
\usepackage{multirow}

\usepackage{amsthm}
\newtheorem{theorem}{Theorem}[section]

\newtheorem{example}[theorem]{Example}

\newcommand{\R}{\mathbb{R}}

\newcommand{\parabreak}{\vspace*{0.1in}}

\usepackage[boxed,vlined,ruled,linesnumbered]{algorithm2e}

\usepackage{fancyheadings}

\title{\textbf{An implementation of CAD in Maple} \\ \textbf{utilising McCallum projection}}
\author{Matthew England}
\date{Department of Computer Science, University of Bath, Bath, UK \\ \texttt{M.England@bath.ac.uk} }

\begin{document}

\maketitle

\pagestyle{fancy}
\lhead{M.~England}
\rhead{An implementation of CAD in Maple utilising McCallum projection}
\cfoot{}

\begin{abstract}
Cylindrical algebraic decomposition (CAD) is an important tool for the investigation of semi-algebraic sets.  Originally introduced by Collins in the 1970s for use in quantifier elimination it has since found numerous applications within algebraic geometry and beyond.  Following from his original work in 1988, McCallum presented an improved algorithm, \texttt{CADW}, which offered a huge increase in the practical utility of CAD.  

In 2009 a team based at the University of Western Ontario presented a new and quite separate algorithm for CAD, which was implemented and included in the computer algebra system {\sc Maple}.  As part of a wider project at Bath investigating CAD and its applications, Collins and McCallum's CAD algorithms have been implemented in {\sc Maple}.  This report details these implementations and compares them to \textsc{Qepcad} and the Ontario algorithm.  

The implementations were originally undertaken to facilitate research into the connections between the algorithms.  However, the ability of the code to guarantee order-invariant output has led to its use in new research on CADs which are minimal for certain problems.  In addition, the implementation described here is of interest as the only full implementation of \texttt{CADW}, (since {\sc Qepcad} does not currently make use of McCallum's delineating polynomials), and hence can solve problems not admissible to other CAD implementations.
\end{abstract}

\noindent This work is supported by EPSRC grant EP/J003247/1.

\thispagestyle{empty}

\section{Introduction} \label{SEC_Intro}

Cylindrical algebraic decomposition (CAD) was first announced in 1973 by Collins.  A CAD is a decomposition of $\R^n$ into cells, constructed with respect to a set of input polynomials in $n$ variables. Each cell can be described as a semi-algebraic set and the cells are cylindrically arranged, meaning the projection of any two cells is either equal or disjoint.  Usually, the CAD produced is such that each polynomial is sign-invariant within each cell, allowing for the solution of many problems defined by the polynomials.  
Collins provided the first algorithm to compute a CAD \cite{Collins75, ACM84I}, developed as a tool for quantifier elimination in real closed fields.  
Since their discovery they have found numerous applications ranging from robot-motion planning \cite{SS83II} to simplification technology \cite[etc.]{BD02, DBEW12}

Collins' algorithm has two phases.  The first, \textit{projection}, applies a projection operator repeatedly to a set of polynomials, each time producing another set of polynomials in one fewer variables.  Together these sets contain the {\em projection polynomials}.  The second phase, \textit{lifting}, then builds the CAD incrementally from these polynomials.  First $\R$ is decomposed into cells which are points and intervals corresponding to the real roots of the univariate polynomials.  Then $\R^2$ is decomposed by repeating the process over each cell using the bivariate polynomials at a sample point of the cell.  The output for each cell consists of {\em sections} of polynomials (where a polynomial vanishes) and {\em sectors} (the regions between these). Together these form the  {\em stack} over the cell, and taking the union of these stacks gives the CAD of $\R^2$.  This process is repeated until a CAD of $\R^n$ is produced.  

To conclude that the CAD of $\R^n$ produced using sample points in this way is sign-invariant we need the key concept of delineability.  A polynomial is {\em delineable} in a cell if the portion of its zero set in the cell consists of disjoint sections.  A set of polynomials are {\em delineable} in a cell if each is delineable and further that the sections of different polynomials in the cell are either identical or disjoint.  A projection operator is valid for use in the algorithms if over each cell of a sign-invariant CAD for projection polynomials in $r$ variables, the polynomials in $r+1$ variables are delineable.

The output of a CAD algorithm depends on the ordering of the variables.  In this paper we usually work with polynomials in $\mathbb{Z}[x_1,\ldots,x_n]$ with the variables, ${\bf x}$, listed in ascending order; (so we first project with respect to $x_n$ and so on until we have univariate polynomials in $x_1$).  The \textit{main variable} of a polynomial, ${\rm mvar}(f)$, is the greatest variable present with respect to the ordering. 

Since Collins published the original algorithm there has been much research into improvements.  These include, but are not restricted to; the use of equational constraints \cite[etc.]{McCallum99}, the notion of partial CADs \cite[etc.]{CH91} and \cite[etc.]{Strzebonski00}, ideas on cell adjacency and clustering \cite[etc.]{ACM84II}, ideas on preprocessing the input \cite[etc]{WBD12_GB} and heuristics to pick the best variable ordering \cite[etc.]{DSS04}.  A summary of developments over the first twenty years of CAD was given by \cite{Collins98}. 

A key area of research has been in improving the projection operator used in the first phase of the algorithm.  Collins original operator can produce CADs with many more cells than required for sign invariance of the polynomials.  McCallum proved in \cite{McCallum88} that for most problems a far simpler projection operator could be used.  Then in \cite{McCallum98} McCallum presented an algorithm, named \texttt{CADW}, detailing how and when the improved operator should be used.  A particular feature of the algorithm is that the CAD produced is not just sign invariant with respect to the input polynomials, but order invariant.  Further development of McCallum's algorithm and operator have been presented by Brown in \cite{Brown01, Brown05}.  

McCallum's projection operator is implemented in {\sc Qepcad} \cite{Brown03}; an interactive command-line program written in C using the \texttt{SACLIB} library of computer algebra functions.  The name stands for \textit{quantifier elimination by partial cylindrical algebraic decomposition} but the software can also produce full cads and offers a wide range of options such as the use of equational constraints and 2d plots.  

In 2009 a new approach to producing CADs was presented in \cite{CMXY09}.  Instead of projection and lifting this approach first computes a decomposition of complex space and then refines to real space, making use of the theory of triangular sets and regular chains \cite[etc.]{ALM99}.  The algorithm was implemented in {\sc Maple} and is included in the \texttt{RegularChains} library \cite[etc.]{MorenoMaza99}, distributed with {\sc Maple}.  

\parabreak

This report gives details on a new implementation of CAD via projection and lifting, within {\sc Maple}.  The implementation was originally undertaken to facilitate research into the connections between the two approaches for computing CAD.  However, its ability to guarantee order-invariant output has led the implementation to utility as part of new algorithm in \cite{BDEMW13} offering decompositions which while not sign invariant for their input polynomials, are minimal for certain problems based on them.  In addition, the implementation is of interest as it is the only \textit{full} implementation of \texttt{CADW}, since {\sc Qepcad} can not produce McCallum's delineating polynomials leading to unnecessary warnings about potential failure for certain examples.  (We note that \textsc{SyNRAC} \cite{YA06} also implements CAD via projection and lifting in {\sc Maple}, but this implementation uses Collins' algorithm and so can not provide order-invariance for use in \cite{BDEMW13}.)

\parabreak

In Section \ref{SEC_CADW} we summarize the theory of the Collins and McCallum projection operators algorithm and in Section \ref{SEC_Maple} we describe our implementation in {\sc Maple}.  In Section \ref{SEC_Exper} we describe how this implementation compares with both {\sc Qepcad} and the algorithm from the theory of regular chains.
This report should be accompanied with a file \texttt{ProjectionCAD.mm} containing the code implementing the algorithms in {\sc Maple}.  The file is designed to be read into {\sc Maple} to define a library of commands, but can also be viewed as a text file.  Details on installing and working with the code are given in Subsection \ref{APP_ProjectionCAD}.

\section{Projection operators} \label{SEC_CADW}

We summarize the projection operators of Collins and McCallum, (for more details see \cite{ACM84I} and \cite{McCallum88}).  

Let $f$ and $g$ be real univariate polynomials.  Denote by $\mathrm{deg}(f)$ the degree of $f$, by $f'$ the first derivative of $f$, by $\mathrm{red}^k(f)$ the $k$th reductum of $f$, by $\mathrm{discr}(f)$ the discriminant of $f$, by $\mathrm{res}(f,g)$ the resultant of $f$ and $g$ and by $\mathrm{psc}_j(f,g)$ the $j$th principal subresultant coefficients of $f$ and $g$.  

Let $F$ be a set of univariate polynomials and suppose $f,g \in F$.  Then denote by $\mathrm{coeff}(F)$, the set of all non-zero coefficients of all elements of $F$, by $\mathrm{red}(F)$ the set of all $\mathrm{red}^k(f)$ such that $0 \leq k \leq \mathrm{deg}(f)$, by $\mathrm{psc}(F)$ the set of all $\mathrm{psc}_j(f,g)$ such that $f \neq g$ and $0 \leq j \leq \mathrm{min}(\mathrm{def}(f), \mathrm{deg}(g))$, by $\mathrm{psd}(F)$ the set of all $\mathrm{psc}_j(f,f')$ such that  $0 \leq j \leq \mathrm{deg}(f)-1$, by $\mathrm{discr}(F)$ the set of all $\mathrm{discr}(f)$ and by $\mathrm{res}(F)$ the set of $\mathrm{res}(f,g)$.  We can now define the operators
\begin{eqnarray*}
PROJ(F) &:=& \mathrm{coeff}(F) \cup \mathrm{psd}(\mathrm{red}(F)) \cup \mathrm{psc}(\mathrm{red}(F)), \\
P(F) &:=& \mathrm{coeff}(F) \cup \mathrm{discr}(F) \cup \mathrm{res}(F).
\end{eqnarray*}
Note that $P(F)$ is a subset of $PROJ(F)$.  When dealing with multivariate polynomials we may think of them as univariate polynomials in the main variable and use the definitions as above.  

When Collins first devised CAD he showed that $P$ may be used as the projection operator for problems in 2 variables.  However, $P$ cannot always be used for problems in an arbitrary number of variables and so \cite{Collins75} specified the algorithm using $PROJ$ instead.  In \cite{McCallum88} McCallum proved that $P$ may also be used for problems in 3 variables and then in \cite{McCallum98} McCallum extended his results to explain when they hold for problems in more variables.

Key to McCallum's proofs is the concept of order invariance.  Recall that a polynomial has order of vanishing $k$ at a point, if $k$ is the smallest non-negative integer such that some partial derivative of the polynomial of order $k$ does not vanish at the point.  A CAD is \textbf{order invariant} with respect to a set of polynomials if each polynomial has constant order of vanishing within each cell.  

Let $F$ be a square free basis of polynomials in $r$ variables.  McCallum's key theorem showed that if $P(F)$ is order invariant over a cell of $\R^{r-1}$ then each element of $F$ either vanishes identically or is order invariant in every cell of the stack above.  Hence the theorem can be applied recursively to prove $P$ is valid, so long as the nullification of a polynomial over a lower-dimensional cell does not occur.  

To classify when the operator $P$ may be used, McCallum defined the following condition.  Let $F$ be a set of polynomials in $r$ variables and let $\mathrm{prim}(F)$ and $\mathrm{cont}(F)$ be respectively the set of primitive parts and the set of contents of $F$.  Then $F$ is \textbf{well-oriented} if either $r=1$ or both the following hold:
\begin{enumerate}[(1)]
\item Every $f \in \mathrm{prim}(F)$ has a finite number of nullification points, i.e. if the main variable is $x_k$ then $f(\alpha,x_k)=0$ for at most a finite number of $\alpha \in \R^{k-1}$.
\item The set $P(F) \cup \mathrm{cont}(F)$ is well oriented.
\end{enumerate}
If condition (1) is changed to the stronger condition of \textit{no nullification points} then the polynomials are said to be \textbf{very well oriented}.

For well-oriented sets of polynomials the projection operator $P$ can be used to generate CADs.  This is shown using McCallum's theorem discussed above and, in the case where are a finite number of nullifying points, by replacing the nullified polynomial $f$ with a \textbf{delineating polynomial}; a partial derivative of $f$ whose roots include the $x_k$ coordinate of every point where the order of $f$ differs from the minimal order induced by $\alpha$.  In \cite{Brown05} Brown defined the \textbf{minimal delineating polynomial} as the case where these are the only roots. 

In \cite{McCallum98} McCallum gave an algorithm \texttt{CADW} which aims to build a CAD using $P$.  Before lifting over each cell the algorithm checks for nullification over the cell.  If the dimension of the cell is zero then a delineating polynomial is used instead and otherwise the algorithm fails, declaring the polynomials not well-oriented.  If the algorithm finishes then the CAD produced is order-invariant.

{\sc Qepcad} offers both McCallum's operator and Hong's modification of Collins' \cite{CH91}.  However, {\sc Qepcad} does not implement the calculation of delineating polynomials and so can fail unnecessarily when using McCallum's operator in cases where the polynomials are well oriented but not very well-oriented.
\footnote{Actually {\sc Qepcad} does not necessarily \textit{fail}, but it will produce a warning message indicating that the output may not be correct.}

However, this situation is minimised since before declaring failure {\sc Qepcad} ensures this is really necessary by running through some checks detailed in \cite{Brown05}.  For example, it may be the case that nullification does not lead to order-invariance, (meaning the minimal delineating polynomial is a constant).  Consider $f=zy-x$ which is nullified by the point $(x,y)=(0,0)$ and so is well oriented but not very well oriented.  However, for all values of $z$ this polynomial has order 1 and so no delineating polynomial need be added, (and {\sc Qepcad} will not fail).  Other checks performed by {\sc Qepcad} include; seeing if a delineating polynomial is already included in the projection set, checking whether the nullified polynomial is one that must actually be order-invariant to ascertain correctness, and checking whether it is the final lift in which case only sign-invariance is required.  

These checks rule out many cases where we may expect {\sc Qepcad} to fail but the example constructed below demonstrates that there are still cases where delineating polynomials are really required to guarantee correct output using McCallum's operator.  

\begin{example}
To construct the example we need a polynomial in $k$ variables nullified by a cell in $k-1$ variables.  We must have $k > 2$ or the nullification will be avoided when considering the content and primitive part individually.  A simple choice is $f=zy-x^2$ which is nullified over a cell with $x=y=0$, which would certainly be produced from its coefficients belonging to the projection set.  Further, this $f$ has non-constant minimal delineating polynomial, $z$.  

However, if we were to just construct a CAD for $f$ then the nullification would occur at the final lift and so we need not bother with a delineating polynomial.  We must instead consider a polynomial $p$ which produces $f$ as a projection factor.  If we let $f$ be a coefficient then we need only make $f$ order-invariant \cite{Brown01} so instead we choose a polynomial for which $f$ is the discriminant: $p=f+w^2$.  

Hence, when considering a CAD for $p$ with respect to the variable ordering $w > z > y > x$, a delineating polynomial will certainly be required.  This example was considered in the experiments detailed in Section \ref{SEC_Exper} and using McCallum's operator a CAD with 73 cells was produced.
\end{example}

\section{Implementation in Maple} \label{SEC_Maple}

Algorithms to construct CADs with both Collins' and McCallum's projection operators have been implemented in {\sc Maple}, with pseudo code for the main algorithms presented below and the actual code freely available from the author's website.  The package is titled \texttt{ProjectionCAD} and is designed to complement the alternative approach of constructing CADs via regular chains which is already distributed with {\sc Maple}.  Indeed, the code described here makes use of tools from the \texttt{RegularChains} package and gives output in the same format.  

Algorithm \ref{alg_CADProjection} computes a set of projection polynomials (including the input polynomials) with respect to the chosen operator.  Algorithm \ref{alg_CADLifting} will use those polynomials and the projection choice to build a CAD of $\R^n$ over which they are sign invariant.  Algorithm \ref{alg_CADFull} is a composition of the two which can generate a variety of CADs via projection and lifting.  

\begin{algorithm} \caption{CADProjection} \label{alg_CADProjection}
\DontPrintSemicolon
\SetKwInOut{Input}{Input}\SetKwInOut{Output}{Output}
\Input{$\bullet$ The input set of polynomials $F \subset \R[x_n,\ldots,x_1]$. \\
$\,\, \bullet$ A choice $proj$ of either Collins or McCallum.
}
\Output{$\bullet$ A set of projection polynomials $P \subset \R[x_n,\ldots,x_1]$.}
\BlankLine
Set $P_1$ to be the finest square-free basis for the primitive parts of $F$\;
Set $cont$ to be the set of contents of $F$.  \;
\For{$i=2,\ldots,n$}{
Set $\hat{P}_{i}$ to be $proj(P_{i}) \cup cont$ excluding any constant polynomials. \; 
Set $P_i$ to be the finest square-free basis for the primitive parts of $\hat{P}_i$. \;
Reset $\textrm{cont}$ to be the set of contents of $\hat{P}_i$. \;  
}
$P := \bigcup_{i=1}^n P_i$\;
\Return $P$\;
\end{algorithm}

Both Collins and McCallum note in their work that for some examples simplifications could be made to the coefficients / reducta included in their operators.  For example, if a polynomial has a constant coefficient then this clearly does not need to be added to the projection polynomials.  Further, no subsequent polynomials would need to be added since there will be no situations where these become leading coefficient.  Some such elementary simplifications have been incorporated in the {\sc Maple} implementation of Algorithm \ref{alg_CADProjection}.

We note that Algorithm \ref{alg_CADLifting} differs depending on the choice of projection operator in Algorithm \ref{alg_CADProjection}, due to the requirements to check that the input polynomials are well-oriented when using McCallum's operator.  There is a further optional input in Algorithm \ref{alg_CADLifting} which can be used to request that the CAD outputted is not only sign-invariant but also order invariant.  If not set to true then the algorithm does not check for nullification on the final lift since without this the CAD of $\R^n$ is still guaranteed to be sign-invariant, sufficient for most applications, (but insufficient for the application in \cite{BDEMW13}).

\begin{algorithm} \caption{CADLifting} \label{alg_CADLifting}
\DontPrintSemicolon
\SetKwInOut{Input}{Input}\SetKwInOut{Output}{Output}
\Input{$\bullet$ A set of polynomials $P \subset \R[x_n,\ldots,x_1]$ from \texttt{CADProjection}. \\
$\,\, \bullet$ A choice $proj$ of either Collins or McCallum. \\
$\,\, \bullet$ A boolean $fcad$ if the final CAD is to be order invariant.
}
\Output{A sign-invariant CAD of $\R^n$, also order-invariant if $fcad=true$.}
\BlankLine
\For{$i=1,\ldots,n$}{
$P_i:= \{ p \in P \mbox{ such that the main variable of p is } x_i \}$. \;
}
Set $C_1$ to be a CAD of $\R$ formed by the decomposition of the real line according to the real roots of the polynomials in $P_n$.\;
\For{$i=2,\ldots,n$}{
\For{\emph{each cell} $c \in C_{i-1}$}{
\eIf{$proj=McCallum$}{
\eIf{$i<n$ \textbf{\textrm{or}} $fcad=true$}{
set $Q_i$ to be the empty set. \;
\For{\emph{each polynomial} $p \in P_{n+1-i}$}{
\eIf{$p=0$ \emph{throughout} $c$}{ 
\eIf{dim(c)=0}{
add the minimal delineating polynomial to $Q_i$ if it is non-constant.\;
}{give warning message about potential failure.}
}{add $p$ to $Q_i$}
}
}{Set $Q_i:=P_i$}
}{Set $Q_i=P_i$}
Set $S_c := \texttt{CADGenerateStack}(c, Q_i)$\;  
}
Set $C_i := \bigcup_c S_c$ \;
}
\Return $C_n$\;
\end{algorithm}

\begin{algorithm} \caption{CADFull} \label{alg_CADFull}
\DontPrintSemicolon
\SetKwInOut{Input}{Input}\SetKwInOut{Output}{Output}
\Input{$\bullet$ The input set of polynomials $F \subset \R[x_1,\ldots,x_n]$. \\
$\,\, \bullet$ A choice $proj$ of either Collins or McCallum.\\
$\,\, \bullet$ A boolean $fcad$ if the final CAD needs to be order invariant. 
}
\Output{$\mathcal{C}$, a CAD of $\R^n$.}
\BlankLine
$P := \texttt{CADProjection}( F, proj )$\;
$C := \texttt{CADLift}( P, proj, fcad )$\;
\Return $C$\;
\end{algorithm}

Algorithm \ref{alg_CADGenerateStack} is a sub-algorithm which describes how the stack generation is implemented.  To generate a stack the real roots of those projection polynomials with main variable $x_i$ need to be calculated when the other variables are set according to a cell of dimension $i-1$.  New cells of dimension $i$ are then defined with the variable $x_i$ set in turn to be those roots and the intervals between.  The roots may not be rational but algebraic numbers, and so care needs to be taken in implementing this.  We make use of the \texttt{RegularChains} library and in particular and internal command for stack generation described in Section 5.2 of \cite{CMXY09}.
Algorithm \ref{alg_CADGenerateStack} ensures that the polynomials passed to the \texttt{RegularChains} algorithm satisfy the assumptions that algorithm makes, namely that the polynomials are co-prime and square-free when evaluated on the cell, (\textit{separate above the cell} in the language of regular chains).  

\begin{algorithm} \caption{CADGenerateStack} \label{alg_CADGenerateStack}
\DontPrintSemicolon
\SetKwInOut{Input}{Input}\SetKwInOut{Output}{Output}
\Input{$\bullet$ A cell $c$ from a CAD of $\R^{i-1}$, or $\emptyset$ if $i=0$.  \\
$\,\, \bullet$ A set of projection polynomials $P \subset \R[x_{n-i},\ldots,x_n]$. \\
}
\Output{$\mathcal{S}$, a stack over $c$ with respect to $P$.}
\BlankLine
Set $\hat{P}$  to be a set of polynomials with the same zeros as $P$ but which are coprime and square-free within the cell $c$.  \;
\tcp{\textrm{By encoding the variables which are set to a point as a regular chain, and using commands from the \texttt{RegularChains} package \cite{MorenoMaza99}.}}
$S := \texttt{RegularChains-GenerateStack}(c,\hat{P})$\;
\Return $S$\;
\end{algorithm}

\subsection{Working with the \texttt{ProjectionCAD} library in Maple} \label{APP_ProjectionCAD}

This report should be accompanied by a file \texttt{ProjectionCAD.mm} containing the code implementing the algorithms in {\sc Maple}.  The file is designed to be read into {\sc Maple} to define a library of commands, but can also be viewed as a text file.  To read the package into {\sc Maple} and make the commands available use
\begin{verbatim}
> read("ProjectionCAD.mm"):
> with(ProjectionCAD):
\end{verbatim}
Commands implementing Algorithms \ref{alg_CADProjection}$-$\ref{alg_CADFull} are now available, with the same names.  For each there are two required arguments; a list of polynomials and a list of variables in descending order.  There are also optional keyword arguments; \texttt{method} to choose between Collins or McCallum, \texttt{output} to specify how the CAD is represented and \texttt{finalOI} to specify that the outputted CAD be order invariant.  Details on the progress of the algorithm can be accessed by setting the \texttt{infolevel} for each command.  

Every cell in a cad of $\R^n$ is equipped with at least an index and a sample point.  The index is an $n$-tuple of integers indicating the value or range of each variable using the real roots in increasing order computed for each stack.  
(Hence a cell whose index consists of only even integers defines a single point of $\R^n$ while an index of only odd integers is a cell of full dimension $n$.)  
The sample points produced by \texttt{ProjectionCAD} are encoded as algebraic numbers given by regular chains and bounds isolating a single root.  This is a consequence of using the internal \texttt{RegularChains} algorithm and is desired as it makes comparing CADs produced by the two approaches simple.

The implementation in {\sc Maple} can also produce cell representations (bounds or values for the variables using algebraic numbers) and if requested (with the \texttt{output} argument) can display the output intuitively using the \texttt{piecewise} construct in {\sc Maple}.  
The number of cells in the CAD may be measured using the \texttt{nops} command on the output, (or the \texttt{CADNumCellsInPiecewise} command if the piecewise output format is used).  

\noindent A simple example of using the code is given below.  The output is as displayed in a {\sc Maple} worksheet, except that the sample points have been replaced by $SP$ for brevity.  
\begin{verbatim}
> f := x^2+y^2-1:
\end{verbatim}
\begin{verbatim}
> cad := CADFull([f], vars, method=McCallum, output=piecewise);
\end{verbatim}
\[
\begin{cases}
SP & \quad x<-1 \\
\begin{cases} SP & \quad y<0 \\ SP & \quad y=0 \\ SP & \quad 0<y \end{cases}  & \quad x=-1 \\
\begin{cases} 
SP & \qquad\qquad y<-\sqrt{-{x}^{2}+1} \\ 
SP & \qquad\qquad y=-\sqrt {-{x}^{2}+1} \\ 
SP & {\it And} \left( -\sqrt {-{x}^{2}+1}<y,y<\sqrt {-{x}^{2}+1} \right) \\ 
SP & \qquad\qquad y=+\sqrt {-{x}^{2}+1} \\ 
SP & \qquad\qquad \sqrt {-{x}^{2}+1}<y 
\end{cases}
&{\it And} \left( -1<x,x<1 \right) \\
\begin{cases} SP & \quad y<0 \\ SP & \quad y=0 \\ SP & \quad 0<y \end{cases}  & \quad x=1  \\
SP & \quad 1<x
\end{cases}
\]
\begin{verbatim}
> CADNumCellsInPiecewise(cad);
\end{verbatim}
\[
13
\]

\section{Experimental results} \label{SEC_Exper}

We have run experiments comparing three of the implementations of CAD.

All the tests were run on a Linux desktop with Intel core i5 CPU (1.6GHz) and 8.0Gb total memory.  The timings in {\sc Maple} were recorded using the inbuilt \texttt{time} function while the timings for {\sc Qepcad} are the total system time recorded at the end of each session.  For the tests in {\sc Maple} we use the \texttt{timelimit} function to limit each example to 1000 seconds 
\footnote{The {\sc Maple} \texttt{timelimit} of 1000s does not apply to kernel operations so it is possible to have timings that are slightly higher, as is the case for the simplified Putnam example in Table \ref{tab_Experiments}.}
of computation time.
For the tests in {\sc Qepcad} we run with the option +N500000000 + L200000, where the first option specifies the memory to be pre-allocated and the second the number of prime numbers to be used.  

The examples for the experiments were all taken from the CAD example repository described in \cite{WBD12_EX} and stored at \texttt{http://opus.bath.ac.uk/29503} and .  Some of the problems here are quantified, however these experiments are designed to compare the CAD implementations only, and so were run on the unquantified versions, (i.e. full sign-invariant CADs were computed for the polynomials involved in the problems).  

\noindent The three implementations compared were as follows.
\begin{description}
\item[PCAD]  We let PCAD denote the implementation of CAD via projection and lifting which is the topic of this paper.  Run in {\sc Maple} 16 through the command \texttt{CADFull} from the \texttt{ProjectionCAD} package developed at Bath.  We run with the default settings; using McCallum projection and worrying about nullification at all except the final lift where this is ignored.
\item[TCAD]  We let TCAD denote the implementation of CAD via triangular sets.  Run in {\sc Maple} 16 with the in-built command from the \texttt{RegularChains} package, 
\texttt{CylindricalAlgebraicDecompose}. 
\item[QCAD] We let QCAD denote the implementation of CAD via projection and lifting given by {\sc Qepcad-B} version 1.69.  The code is given the \texttt{full-cad} option and default settings otherwise; McCallum projection and lifting with various improvements \cite{Brown01, Brown05} but no delineating polynomials.
\end{description}
We note that there are other implementations of CAD available, including Mathematica, Redlog and SyNRAC.

The results of the experiments are displayed in Table \ref{tab_Experiments}.  An F indicates failure for theoretical reasons
\footnote{this includes the production of warnings that the output may not be correct}
while a T/O indicates failure due to timeout in {\sc Maple}.  We note that with the setting described above, {\sc Qepcad} never failed due to timeout or memory issues.  

Unless it fails for theoretical reasons, QCAD is usually the quickest.  The examples where it was slower were generally small examples in which the time for QCAD was dominated by its initialisation.  It is likely that {\sc Qepcad} still represents the state of the art for computing full CADs at the moment.  However, we make the cautionary note that the algorithm for TCAD described in \cite{CMXY09} is to be replaced by a significantly more efficient version, designed to reduce repeated computations and perform some steps in an incremental manner.  This improvement is described in the preprint \cite{CM12_CAD_Arxiv}, but the implementation is still under development and not yet available.  

Next we note that TCAD never fails for theoretical reasons and is usually (but not always) quicker than PCAD.  Comparing PCAD with QCAD we see that the cell counts are usually identical, as expected since they implement the same theoretical algorithm.  There are occasions when QCAD will fail when PCAD does not because it does not implement McCallum's delineating polynomials, as in the example from Section \ref{SEC_CADW} and the Quartic example.  However, there are also examples where PCAD failed when QCAD did not, (the Whitney umbrella and the x-axis ellipse problem).  These theoretical failures were avoided in QCAD since they correspond to cases where nullification does not contradict the theory underpinning the algorithm, which {\sc Qepcad} checks for (as discussed in Section \ref{SEC_CADW}).  It is possible to instruct \texttt{ProjectionCAD} to continue the computation following a failure, which for these examples would lead to the same number of cells as QCAD.  The other examples where the cell count differs are due to QCAD using a partial implementation of the simplified operator described in \cite{Brown01}.

\begin{table}[hb]
\centering
\begin{tabular}{|l|cc|cc|cc|}
\hline
\multirow{2}{*}{\textbf{Problem}}    & \multicolumn{2}{|c|}{\textbf{PCAD}} & \multicolumn{2}{|c|}{\textbf{TCAD}} & \multicolumn{2}{|c|}{\textbf{QCAD}} \\
                                     & cells & time   & cells & time   & cells & time  \\
\hline
Parametric parabola                  & 115   & 0.5    & 27    & 0.1    & 115    & 1.6  \\ 
Whitney umbrella                     & F     & -      & 895   & 3.4    & 895    & 3.8  \\ 
Quartic                              & 333   & 3.9    & 233   & 2.7    & F      & -    \\ 
Sphere \& catastrophe                & 509   & 6.2    & 421   & 4.140  & 509    & 3.7  \\ 
Arnon-84                             & 55    & 0.3    & 55    & 0.157  & 55     & 3.7  \\ 
Arnon-84-2                           & 41    & 0.4    & 41    & 0.332  & 41     & 3.7  \\ 
Implicitization                      & F     & -      & 895   & 5.228  & F      & -    \\ 
Ball \& cylinder                     & 365   & 4.2    & 365   & 3.914  & 365    & 3.8  \\ 
Term rewrite system                  & 1099  & 8.2    & 1099  & 7.768  & 1099   & 3.6  \\ 
Collins and Johnson                  & 3673  & 50.2   & 3673  & 65.438 & 3673   & 3.8  \\ 
Lower bounds range                   & F     & -      & 333   & 1.550  & F      & -    \\ 
X-axis ellipse problem               & F     & -      & 20225 & 252.0  & 62645  & 5.1  \\ 
Davenport \& Heintz                  & 4949  & 23.8   & 4949  & 21.0   & 4949   & 3.8  \\ 
Hong-90                              & 27    & 0.1    & 27    & 0.2    & 27     & 3.5  \\ 
Solotareff-3                         & -     & T/O    & -     & T/O    & 243325 & 12.3 \\ 
Collision problem                    & -     & T/O    & -     & T/O    & 45979  & 5.0  \\ 
Random trivariate                    & -     & T/O    & -     & T/O    & 877    & 14.9 \\ 
Off-center ellipse                   & 4569  & 153.9  & 2705  & 51.5   & 4593   & 5.0  \\ 
Concentric circles                   & 41    & 0.2    & 41    & 0.2    & 41     & 3.7  \\ 
Non-concentric circles               & 41    & 0.2    & 41    & 0.3    & 41     & 3.7  \\ 
Simplified ESP                       & -     & T/O    & -     & T/O    & 56105  & 10.2 \\ 
Simplified Putnum                    & 55021 & 1026.0 & 10517 & 193.0  & 10517  & 3.9  \\ 
Simplified YangXia                   & -     & T/O    & -     & T/O    & 6313   & 5.2  \\ 
Simplified SEIT                      & -     & T/O    & -     & T/O    & F      & -    \\ 
Cyclic-3                             & 381   & 3.7    & 381   & 4.2    & 381    & 3.8  \\ 
Example from \S \ref{SEC_CADW}       & 73    & 0.2    & 67    & 0.1    & F      & -    \\ 
\hline
\end{tabular}
\caption{Table detailing timings and cell counts for computation of full CADs using the three implementations described in Section \ref{SEC_Exper}.  }
\label{tab_Experiments}
\end{table}

\section{Summary} \label{SEC_Summary}

We have described an implementation of CAD via projection and lifting in {\sc Maple}.  The implementation offers both McCallum's and Collins' operators with a variety of customizations available.  The implementation was undertaken to facilitate research into the connections between the different approaches to CAD.  The experimental results demonstrate that for most problems {\sc Qepcad} is superior.  Nevertheless, the approach described here offers significant utility for two main reasons.
\begin{enumerate}[(1)]
\item It implements McCallum's delineating polynomials (actually Brown's minimal delineating polynomial) and thus can produce the only full implementation of McCallum's \texttt{CADW}.  We have demonstrated that there are examples in which {\sc Qepcad} will fail for which this implementation can succeed.  
\item It can offer the user the choice of an order-invariant CAD in the final output.  This is essential for the application in \cite{BDEMW13} which defines a new type of CAD, minimal for certain problems.  
\end{enumerate}
Examples have been given where the output of CAD via Regular Chains in {\sc Maple} is not order-invariant and it is not currently known how the algorithm could be modified to produce order-invariance, (further research is currently being undertaken).  

It is likely that a small tweak order the hood in {\sc Qepcad} could be made to offer order-invariant output at in (2).  However, this would then greatly amplify the utility described in (1) since the majority of cases of nullification occur at the final lift where they can be safely ignored if only a sign-invariant CAD is required.  For example, consider again $f=zy-x^2$ from Section \ref{SEC_CADW}.  We noted there that $f$ has non-constant minimal delineating polynomial and so to give an order-invariant CAD of $f$ alone, this would need to be included.  Running PCAD gives a CAD of 21 cells (as does TCAD and QCAD).  One of these cells has $x=0,y=0,x$ free, over which $f$ is not order-invariant.  However, if we run \texttt{CADFull} using the optional argument to specify an order-invariant output then this cell is split and the CAD outputted has 23 cells.

If this implementation remains essential for applications such as \cite{BDEMW13} then some future improvements to the package are likely.  These should include the implementation of some of the improvements described in \cite{Brown05} to avoid failure unless it is absolutely necessary.  Further, an implementation of the Brown-McCallum operator \cite{Brown01} is very desirable to offer the lowest possible cell counts. 

\begin{footnotesize}
\bibliography{CAD}{}

\begin{thebibliography}{10}

\bibitem{ACM84I}
D.~Arnon, G.E. Collins, and S.~McCallum.
\newblock Cylindrical algebraic decomposition {I}: The basic algorithm.
\newblock {\em SIAM J, Comput.}, 13:865--877, 1984.

\bibitem{ACM84II}
D.~Arnon, G.E. Collins, and S.~McCallum.
\newblock Cylindrical algebraic decomposition {II}: An adjacency algorithm for
  the plane.
\newblock {\em SIAM J, Comput.}, 13:878--889, 1984.

\bibitem{ALM99}
P.~Aubry, D.~Lazard, and M.~Moreno Maza.
\newblock On the theories of triangular sets.
\newblock {\em J. Symb. Comput.}, 28(1-2):105--124, 1999.

\bibitem{BD02}
R.~Bradford and J.H. Davenport.
\newblock Towards better simplification of elementary functions.
\newblock In {\em Proceedings of the 2002 international symposium on symbolic
  and algebraic computation ({ISSAC})}. ACM, 2002.

\bibitem{BDEMW13}
R.~Bradford, J.H. Davenport, M.~England, S.~McCallum, and D.~Wilson.
\newblock Cylindrical algebraic decompositions for boolean combinations.
\newblock {\em Submitted for publication}. Preprint at \texttt{http://opus.bath.ac.uk/33926/}, 2013.

\bibitem{Brown01}
C.W. Brown.
\newblock Improved projection for cylindrical algebraic decomposition.
\newblock {\em J. Symb. Comput.}, 32(5):447--465, 2001.

\bibitem{Brown03}
C.W. Brown.
\newblock {QEPCAD B}: {A} program for computing with semi-algebraic sets using
  {CAD}s.
\newblock {\em ACM SIGSAM Bulletin}, 37(4):97--108, 2003.

\bibitem{Brown05}
C.W. Brown.
\newblock The {M}c{C}allum projection, lifting, and order-invariance.
\newblock Technical report, U.S. Naval Academy, Computer Science Department,
  2005.

\bibitem{CM12_CAD_Arxiv}
C.~Chen and M.~Moreno Maza.
\newblock An incremental algorithm for computing cylindrical algebraic
  decompositions.
\newblock {\em Preprint: arXiv:1210.5543v1}, 2012.

\bibitem{CMXY09}
C.~Chen, M.~Moreno Maza, B.~Xia, and L.~Yang.
\newblock Computing cylindrical algebraic decomposition via triangular
  decomposition.
\newblock In {\em Proceedings of the 2009 international symposium on Symbolic
  and algebraic computation ({ISSAC})}, pages 95--102. ACM, 2009.

\bibitem{Collins75}
G.E. Collins.
\newblock Quantifier elimination for real closed fields by cylindrical
  algebraic decomposition.
\newblock In {\em Proceedings of the 2nd GI Conference on Automata Theory and
  Formal Languages}, pages 134--183. Springer-Verlag, 1975.

\bibitem{Collins98}
G.E. Collins.
\newblock Quantifier elimination by cylindrical algebraic decomposition -- 20
  years of progress.
\newblock In B.~Caviness and J.~Johnson, editors, {\em Quantifier Elimination
  and Cylindrical Algebraic Decomposition}, Texts \& Monographs in Symbolic
  Computation, pages 8--23. Springer-Verlag, 1998.

\bibitem{CH91}
G.E. Collins and H.~Hong.
\newblock Partial cylindrical algebraic decomposition for quantifier
  elimination.
\newblock {\em J. Symb. Comput.}, 12:299--328, 1991.

\bibitem{DBEW12}
J.D. Davenport, R.~Bradford, M.~England, and D.~Wilson.
\newblock Program verification in the presence of complex numbers, functions
  with branch cuts etc.
\newblock In {\em 14th International Symposium on Symbolic and Numeric
  Algorithms for Scientific Computing}, SYNASC 2012, 2012.

\bibitem{DSS04}
A.~Dolzmann, A.~Seidl, and T.~Sturm.
\newblock Efficient projection orders for {CAD}.
\newblock In {\em Proceedings of the 2004 international symposium on Symbolic
  and algebraic computation}, ISSAC 2004, pages 111--118. ACM, 2004.

\bibitem{MorenoMaza99}
M.~Moreno Maza.
\newblock On triangular decompositions of algebraic varieties.
\newblock Technical report, NAG Technical Report, 1999.

\bibitem{McCallum88}
S.~McCallum.
\newblock An improved projection operation for cylindrical algebraic
  decomposition of three-dimensional space.
\newblock {\em J. Symb. Comput.}, 5(1-2):141--161, 1988.

\bibitem{McCallum98}
S.~McCallum.
\newblock An improved projection operation for cylindrical algebraic
  decomposition.
\newblock In B.~Caviness and J.~Johnson, editors, {\em Quantifier Elimination
  and Cylindrical Algebraic Decomposition}, Texts \& Monographs in Symbolic
  Computation, pages 242--268. Springer-Verlag, 1998.

\bibitem{McCallum99}
S.~McCallum.
\newblock On projection in {CAD}-based quantifier elimination with equational
  constraint.
\newblock In {\em Proceedings of the 1999 international symposium on Symbolic
  and algebraic computation}, ISSAC '99, pages 145--149. ACM, 1999.

\bibitem{SS83II}
J.T. Schwartz and M.~Sharir.
\newblock On the ``{P}iano-{M}overs'' {P}roblem: {II.} {G}eneral techniques for
  computing topological properties of real algebraic manifolds.
\newblock {\em Adv. Appl. Math.}, 4:298--351, 1983.

\bibitem{Strzebonski00}
A.~Strzebo\'{n}ski.
\newblock Solving systems of strict polynomial inequalities.
\newblock {\em J. Symb. Comput.}, 29(3):471--480, 2000.

\bibitem{WBD12_EX}
D.J. Wilson, R.J. Bradford, and J.H. Davenport.
\newblock A repository for {CAD} examples.
\newblock {\em ACM Communications in Computer Algebra}, 46(3):67--69, 2012.

\bibitem{WBD12_GB}
D.J. Wilson, R.J. Bradford, and J.H. Davenport.
\newblock Speeding up cylindrical algebraic decomposition by {G}r\"{o}bner
  bases.
\newblock {\em Lecture Notes in Computer Science}, 7362:280--294, 2012.

\bibitem{YA06}
H.~Yanami and H.~Anai.
\newblock Development of {SyNRAC}.
\newblock In {\em Proceedings of the 6th international conference on
  Computational Science: Part II. (LNCS vol 3992)}, ICCS '06, pages 462--469,
  2006.

\end{thebibliography}
\bibliographystyle{plain}
\end{footnotesize}

\end{document}